\documentclass[conference]{IEEEtran}
\IEEEoverridecommandlockouts
\usepackage[T1]{fontenc}
\usepackage{graphicx}
\usepackage{subcaption}
\usepackage{amsmath, amssymb}
\usepackage{siunitx}
\usepackage{tikz}
\usepackage{tikzscale}
\usepackage{pgfplots}
\usetikzlibrary{patterns}
\usetikzlibrary{calc}
\usetikzlibrary{snakes,arrows,shapes}

\pgfplotsset{compat=1.18}

\usepackage[hidelinks]{hyperref}
\usepackage{color}

\begin{document}

\title{Evaluating Rapid Makespan Predictions for Heterogeneous Systems with Programmable Logic}

\author{\IEEEauthorblockN{1\textsuperscript{st} Martin Wilhelm}
\IEEEauthorblockA{
\textit{Otto-von-Guericke University}\\
Magdeburg, Germany \\
martin.wilhelm@ovgu.de}
\and
\IEEEauthorblockN{2\textsuperscript{nd} Franz Freitag}
\IEEEauthorblockA{
\textit{University of Applied Sciences}\\
Magdeburg, Germany \\
franz.freitag@h2.de}
\and
\IEEEauthorblockN{3\textsuperscript{rd} Max Tzschoppe}
\IEEEauthorblockA{
\textit{Otto-von-Guericke University}\\
Magdeburg, Germany \\
max.tzschoppe@ovgu.de}
\and
\IEEEauthorblockN{4\textsuperscript{th} Thilo Pionteck}
\IEEEauthorblockA{
\textit{Otto-von-Guericke University}\\
Magdeburg, Germany \\
thilo.pionteck@ovgu.de}
}
\maketitle
\begin{abstract}

Heterogeneous computing systems, which combine general-purpose processors with specialized accelerators, are increasingly important for optimizing the performance of modern applications. A central challenge is to decide which parts of an application should be executed on which accelerator or, more generally, how to map the tasks of an application to available devices. Predicting the impact of a change in a task mapping on the overall makespan is non-trivial. While there are very capable simulators, these generally require a full implementation of the tasks in question, which is particularly time-intensive for programmable logic. A promising alternative is to use a purely analytical function, which allows for very fast predictions, but abstracts significantly from reality. Bridging the gap between theory and practice poses a significant challenge to algorithm developers.
This paper aims to aid in the development of rapid makespan prediction algorithms by providing a highly flexible evaluation framework for heterogeneous systems consisting of CPUs, GPUs and FPGAs, which is capable of collecting real-world makespan results based on abstract task graph descriptions. We analyze to what extent actual makespans can be predicted by existing analytical approaches. Furthermore, we present common challenges that arise from high-level characteristics such as data transfer overhead and device congestion in heterogeneous systems.
\end{abstract}

\begin{IEEEkeywords}
Makespan Prediction, Heterogeneous Systems, Modeling, FPGA, Task Mapping, Design Space Exploration.
\end{IEEEkeywords}

\section{Introduction}

A central challenge in the design of modern high-performance computing systems is the distribution of tasks across various types of specialized processing devices, such as GPUs, FPGAs, or AI units. This process is commonly referred to as \emph{task mapping}. Frequently, this task distribution is done manually by an experienced designer. However, in highly heterogeneous systems with many different tasks and processing devices, it becomes increasingly complex to predict whether the chosen mapping is optimal or even whether it is better than a simple CPU mapping.
This problem intensifies when programmable logic is used to accelerate a design. Even with the use of high-level synthesis (HLS), designing programmable logic is highly time-consuming. Hence, iterations of the design are very expensive and, therefore, undesirable. Instead, avoiding suboptimal guesses early on can strongly reduce the overall cost of the design.

Ideally, a beneficial mapping can be found before an actual implementation takes place. For this, a makespan prediction algorithm for task mapping should be able to find a reliable estimate based on a high-level task description. Current approaches for predicting the runtime of an FPGA implementation typically assume that the FPGA has already been chosen for the specific task and instead focus on aiding the designer in the optimization of the kernel. Therefore, they usually require an existing HLS or annotated C implementation~\cite{oneal2018}. For this reason, they are not suitable for evaluating a high number of potential task mapping scenarios at an early design stage.

In order to make an educated decision about a task mapping, the level of precision provided by prediction algorithms that use implementation data is not needed. Instead, a rough estimate, which can be computed in rapid succession, is generally better suited to decide whether an implementation should be pursued or not. In particular, even with an accurate estimation for the execution time of a single task on all available devices, it is often hard to decide which device should be used due to data transfer times and device congestion. Thus, for complex systems, this process should be automated using a task mapping algorithm. If the prediction can be done multiple times on the fly during the execution of a mapping algorithm, this can drastically improve the results of the optimization.

A common critique of purely analytical prediction algorithms is that they rely on an abstract model, which may not sufficiently reflect the behavior of the real hardware. However, conducting a practical evaluation of whether a model can represent all main characteristics of a complex system is notoriously difficult. Designing and implementing a complex application scenario that involves various devices and a large number of interdependent tasks is extremely time-consuming. Yet, in order to reliably assess the strength of a model, the model must be tested against hundreds of such systems, which is not feasible in practice.

The overall goal of this paper is to aid in the creation of rapid makespan prediction algorithms by providing a structured way to evaluate such predictions in a large number of different application scenarios. In the context of this paper, we focus on the prediction of the final makespan of the heterogeneous system based on pre-evaluated task characteristics. We present an OpenCL-based framework, which can generate large numbers of random test applications in an environment consisting of one or more CPUs, GPUs and FPGAs. With this, we aim to give directions where rough estimates are reasonable and discuss common obstacles for such prediction algorithms.
The evaluation framework is publicly available on GitHub~\cite{taskmappingevaluator}. We expect that the process of evaluating and adapting makespan prediction algorithms using our framework will improve the quality of predictions considerably and increase the trust in the underlying model.

In Section~\ref{sec:sota}, we provide a short overview on the relevant state of the art with a focus on modeling approaches for the task mapping problem. In Section~\ref{sec:framework}, we describe the evaluation framework in detail. In Section~\ref{sec:evaluation}, we use the evaluation framework to generate random sample applications. We then incorporate a rapid makespan prediction into a simulated annealing algorithm in order to compute beneficial task mappings for the sample applications. Finally, we compare the actual resulting makespan on a real system with the predicted makespan.

\section{State of the Art}
\label{sec:sota}

In this section, we give a short overview on existing approaches on task mapping and makespan prediction for heterogeneous systems.

\subsection{Task Mapping}

The mapping of tasks to devices poses a central challenge in the design of heterogeneous systems. Depending on the focus, the problem is closely related to heterogeneous task scheduling and workload partitioning / workload placement. Task mapping approaches can generally be divided into static and dynamic approaches, where static approaches try to find an optimal mapping at design time and dynamic approaches can change the mapping based on runtime information~\cite{carvalho2024,mittal2015}. A rapid makespan prediction can be incorporated in both static and dynamic algorithms. However, if FPGAs are involved, the mapping needs to be at least partially static in order to account for the high design cost of programmable logic. Dynamic task mapping is often used in applications with many recurring tasks and few dependencies. In contrast, for static task mapping, the application is usually modeled using a task graph, i.e., a directed acyclic graph containing all (sub)tasks of the application and the dependencies between them.

Finding a makespan-optimal static task mapping of a task graph is NP-hard, even if the execution times of all tasks could be perfectly predicted on each device~\cite{baruah2004}. Static task mapping approaches frequently employ generalized optimization algorithms in order to tackle this problem. Common approaches are integer linear programs~\cite{emeretlis2022,mohammadi2023,wilhelm2023,zhou2014}, metaheuristics such as simulated annealing~\cite{orsila2007,orsila2013} or genetic algorithms~\cite{alexandrescu2015,erbas2006}, and machine learning~\cite{grewe2011,grewe2013}.

All of these algorithms rely heavily on predefined cost models, which are reflected in their respective optimization criteria. However, most of these approaches use very simplified cost functions to predict the makespan of a given mapping, such as the maximum of the sum of execution times over all devices.
These simple cost functions produce reasonably good results in basic environments but reach their limits in scenarios where data transfer costs are significant, tasks have a high number of dependencies or dataflow streaming between tasks is possible. Exchanging the cost function of these algorithms by a strong and fast makespan prediction algorithm can significantly increase the result quality. Recently, it was shown that genetic algorithms are particularly well suited to benefit from a better makespan prediction in a complex scenario~\cite{wilhelm2025seriesparallel}. Furthermore, in the same work, the authors demonstrate that an efficient cost function enables the design of problem-specific heuristics for task mapping, which are tailored to the individual characteristics of the problem and can outperform generalized algorithms in many use cases.

Generalized task mapping algorithms that explicitly recognize the challenges and opportunities of dataflow processors are still rare. Most work on mapping approaches for FPGAs comes from the field of hardware/software partitioning~\cite{mhadhbi2016}. Here, FPGAs or ASICs are usually viewed as accelerators for selected functions of a software application instead of viewing them as parts of a heterogeneous system consisting of many devices. In consequence, software processors such as GPUs are often neglected in the analysis, and hence, the holistic perspective on the makespan in a heterogeneous system is not taken.
Notable examples for this approach are tools such as MPSeeker~\cite{zhong2017} and AccelSeeker~\cite{zacharopoulos2019}, which try to find profitable tasks to be accelerated based on a high-level software description.

\subsection{Makespan prediction}

Predicting the makespan of an application mapped onto a heterogeneous system consists of two fundamental aspects. First, the prediction of the execution time of individual tasks on their assigned device and, second, the prediction of the overall makespan based on the individual execution times.

Various work exists on the prediction of individual execution times, ranging from hand-tuned analytical models to fully automated code analysis and cycle-accurate simulation~\cite{oneal2018}. For FPGAs, there are various highly accurate models, which can yield an expected execution time in significantly less time than the bitstream generation takes. Still, these models require HLS kernels~\cite{choi2017,koeplinger2016,zhao2017} or annotated C code~\cite{zhong2017}, which might not be feasible to create for a large number of potential tasks.

There is little work that focuses on predicting the overall makespan in a heterogeneous system containing programmable logic. Dreimann et al.\ provide a simulation framework in the context of scheduling that considers CPUs, GPUs and FPGAs using randomly generated execution times~\cite{dreimann2024}.
An evaluation algorithm that also considers dataflow streaming between tasks is presented by Wilhelm et al.~\cite{wilhelm2023}. The authors use a platform model based on basic hardware characteristics in order to compute execution times for their randomly generated tasks. The application model consists of a task graph with tasks connected by edges representing data dependencies. The makespan is predicted in three steps. First, the tasks and edges are topologically sorted. Second, tasks mapped on FPGAs that allow streaming are grouped together into supernodes. Third, the finish times of all processors are updated according to the computed topological order by increasing the time for each task and synchronizing time between devices if data transfers are necessary. For the demonstration of our framework, we use the framework to analyze the behavior of this makespan prediction algorithm.

\section{Framework}
\label{sec:framework}

We propose an OpenCL-based framework for the analysis of makespan predictions for given task mappings. The framework allows for the creation of a large number of diverse test applications in the form of annotated task graphs, which are accompanied by OpenCL kernels that reflect the individual characteristics of the created tasks. The generated kernels can then be employed to build a real heterogeneous system and collect real-world runtime data. This approach enables a fast and efficient determination of the real makespan, which, in turn, can be compared with a previously computed makespan prediction.

In our approach, we focus on assuring the relative validity of the overall makespan prediction. For this, it is not imperative that the execution time of a single task is predicted with high accuracy (as long as it stays in reasonable bounds), but that beneficial mappings can be correctly identified. Hence, we focus on reflecting core characteristics of CPUs, GPUs and FPGAs in the generated OpenCL test kernels.
In the following two subsections, we provide a more detailed overview of the structure of the framework, with a particular focus on the design of the FPGA kernel.

\subsection{Structure}
\label{ssc:structure}

The presented framework consists of three main components. The \emph{application generator} builds random task graphs, annotates them based on key characteristics extracted from the available hardware and assigns a mapping to the graph using one or more provided mapping algorithms. The \emph{kernel generator} takes annotated task graphs together with their mapping and generates appropriate kernels, configuration files and make files. It generates pure OpenCL kernels for tasks mapped on a CPU or GPU and Vitis HLS kernels for tasks mapped on the FPGA. Lastly, during \emph{heterogeneous integration and evaluation}, the OpenCL kernels and the bitstream of the Vitis HLS kernels are incorporated into one heterogeneous system and the actual makespan of the application represented by the mapped task graph is determined.

\begin{figure}[htb]
    \centering
   \resizebox{\linewidth}{!}{\begin{tikzpicture}
    \node [draw, anchor = north west, inner sep = 0mm, minimum width = 4cm, minimum height = 5mm] (TG) at (-4.5,1.75) {Annotated task graph};
    \node [draw, anchor = north west, inner sep = 0mm, minimum width = 4cm, minimum height = 5mm] (HT) at (-4.5,1) {Hardware topology};
    
    \node [draw, anchor = north west, inner sep = 0mm, minimum width = 4cm, minimum height = 5mm] (TMG) at (0,0) {Mapped task graph};

    \node [draw, anchor = north west, inner sep = 0mm, minimum width = 4cm, minimum height = 5mm] (KC) at (0,-1) {Kernel generation};


    \node [draw, fill = white, anchor = north, inner sep = 0mm, minimum width = 1cm, minimum height = 5mm] (CPU) at (-0.2,-2) {};
    \node [draw, fill = white, anchor = north, inner sep = 0mm, minimum width = 1cm, minimum height = 5mm] at (-0.3,-2.1) {};
    \node [draw, fill = white, anchor = north, inner sep = 0mm, minimum width = 1cm, minimum height = 5mm] at (-0.4,-2.2) {};
    \node [anchor = north, inner sep = 1mm] (CPU_1) at (-0.4,-2.8) {CPU/GPU kernels};

    \node [draw, fill = white, anchor = north, inner sep = 0mm, minimum width = 1cm, minimum height = 5mm] (FPGA) at (4.5,-2) {};
    \node [draw, fill = white, anchor = north, inner sep = 0mm, minimum width = 1cm, minimum height = 5mm] at (4.4,-2.1) {};
    \node [draw, fill = white, anchor = north, inner sep = 0mm, minimum width = 1cm, minimum height = 5mm] at (4.3,-2.2) {};
    \node [anchor = north, inner sep = 1mm] (FPGA_1) at (4.3,-2.8) {FPGA kernels};

    \node [draw, anchor = north, inner sep = 0mm, minimum width = 3cm, minimum height = 5mm] (BS) at (4.3,-3.7) {Bitstream};

    \node [draw, anchor = north west, inner sep = 0mm, minimum width = 4cm, minimum height = 5mm] (HKE) at (0,-5) {Host and kernel execution};

    \node [draw, anchor = north west, inner sep = 0mm, minimum width = 4cm, minimum height = 5mm, pattern={north east lines}] (ET) at (0,-6) {\colorbox{white}{Real makespan}};

    \node [inner sep = 0mm] (ADD) at (2, 0.75) {$\bigoplus$};
    \draw [thick, -stealth] (TG) -- (2,1.5) -- (ADD);
    \draw [thick, -stealth] (HT) -- (ADD);
    \draw [thick, -stealth] (ADD) -- (TMG);

    \draw [thick, -stealth] (TMG) -- (KC);

    \draw [thick, -stealth] (KC) -- (CPU);
    \draw [thick, -stealth] (KC) -- (FPGA);

    \draw [thick, -stealth] (CPU_1) -- (HKE);
    \draw [thick, -stealth] (FPGA_1) -- (BS);
    \draw [thick, -stealth] (BS) -- (HKE);

    \draw [thick, -stealth] (HKE) -- (ET);

    \node [inner sep = 0mm, anchor = west] at (2.25, 0.75) {Mapping algorithm};

    \node [inner sep = 1mm, anchor = west] (HLS) at (6, -3.4) {HLS};
    \draw [thick, dashed, -stealth] (HLS) -- (4.4, -3.4);

    \node [draw, anchor = north west, inner sep = 0mm, minimum width = 4cm, minimum height = 5mm] (PA) at (-4.5,-5) {Prediction Algorithm};
    
    \node [draw, anchor = north west, inner sep = 0mm, minimum width = 4cm, minimum height = 5mm, pattern={north east lines}] (P) at (-4.5,-6) {\colorbox{white}{Predicted Makespan}};

    \node [draw, anchor = north west, inner sep = 0mm, minimum width = 4cm, minimum height = 5mm] (C) at (0,-7) {\textbf{Comparison}};

    \draw [thick, -stealth] (ET) -- (C);
    \draw [thick, -stealth] (TMG) -- (-2.5, -0.25) -- (PA);
    \draw [thick, -stealth] (-2.5, -4.65) -- (-0.25, -4.65) -- (-0.25, -5.25) -- (HKE);
    \draw [thick, -stealth] (PA) -- (P);
    \draw [thick, -stealth] (P) -- (-2.5, -7.25) -- (C);
    
    \draw [thick, dotted] (-4.5, -0.75) -- (8.5, -0.75);
    \draw [thick, dotted] (-4.5, -4.5) -- (8.5, -4.5);
    
    \node [anchor = north east] at (8.5,1.75) {\textbf{Application Generator}};
    \node [anchor = north east] at (8.5,-1) {\textbf{Kernel Generator}};
    \node [anchor = north east,align=right] at (8.5,-5) {\textbf{Heterogeneous}\\ \textbf{Integration \& Evaluation}};
    
\end{tikzpicture}}
    \caption{The process flow of the evaluation framework. The framework enables the comparison of a makespan prediction based on an annotated task graph with the real makespan in an equivalent heterogeneous system for arbitrary task graphs.}
    \label{fig:framework_concecpt}
\end{figure}
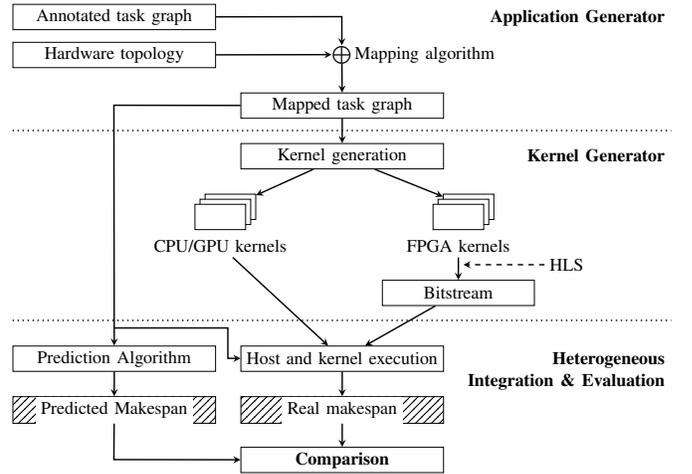

In Figure~\ref{fig:framework_concecpt}, the process flow of the framework is depicted. For the input, the framework supports the generation of random task graphs by the application generator as well as the use of custom task graphs, which allows for a deliberate evaluation of special cases. The graphs are then mapped using a suitable mapping algorithm. After the execution of the host program finishes, the output can be compared against a makespan prediction. Both the mapping and the prediction algorithm are fully exchangeable. In the context of this work, we use a simulated annealing algorithm to determine a suitable mapping and the makespan prediction algorithm of Wilhelm et al.~\cite{wilhelm2023}, which is also used as the optimization criterion for simulated annealing. We furthermore generate pure CPU mappings as a baseline for evaluating the optimization potential.

\subsection{Application generator}

The application generator annotates the tasks of randomly generated task graphs based on three fundamental characteristics: complexity, parallelizability, and streamability.

\begin{itemize}
    \item The \textit{complexity} quantifies the number of operations per data point a task requires, assuming sequential execution. 
    \item The \textit{parallelizability} represents the proportion of the task that can be (arbitrarily well) parallelized.
    \item The \textit{streamability} refers to the maximum gain achievable through a dataflow pipeline, i.e., the ratio between the overall complexity and the complexity of the largest non-divisible subtask.
\end{itemize}

These three characteristics, while strongly simplified, embody the main differences in acceleration potential of the three processing devices considered in this work. A high parallelizability heavily favors the use of a GPU if the task is sufficiently complex. A high streamability favors the implementation on an FPGA. This effect amplifies if data can be streamed between multiple tasks, which is accounted for in the underlying evaluation model and is reflected in the design of the FPGA kernels (Section~\ref{ssc:fpgaarchitecture}).
While a multicore CPU can reasonably well exploit parallelizability, compared to the other two device types, it shines in the execution of complex applications with a large part that must be serially executed and cannot make use of data streaming. Here, the CPU is usually preferable due to its higher clock rate and the fast connection to the main RAM.
For this work, the values for the three parameters are randomly generated. The complexity and streamability of a task are generated using a lognormal distribution with $\mu=2$ and $\sigma=0.5$, whereas a complexity of $1$ represents $100$ actual operations. The parallelizability is set to \SI{100}{\percent} for half of the tasks and to a uniformly distributed value between \SI{0}{\percent} and \SI{100}{\percent} for the rest of the tasks. This choice takes into account that even a small serial part can render GPU executions nearly useless, as it quickly dominates the execution time in a highly parallel environment.
Finally, given the annotated task graph and a suitable mapping algorithm, the application generator assigns processing devices to the given tasks and forwards the mapped task graph to the kernel generator.

\subsection{Kernel generator}
\label{ssc:kernelgenerator}

The kernel generator creates dummy kernels from the annotated task graph that aim to accurately reflect the behavior dictated by the annotations. In essence, all generated kernels take one or more input data arrays $A^{(1)},\dots, A^{(k)}$ of a predefined size $n$ and create a result array $R$ by repeatedly executing the computation $r_i = (r_i + a_i^{(j)}) \mod 256$ for each input array $j\in\lbrace 1,\cdots,k\rbrace$ and each array value $i \in \lbrace 1,\cdots,n\rbrace$. The computation is chosen such that all inputs are processed and the produced values stay in a meaningful range. Since the computation consists of two operations, in order to match the complexity requirements, it is executed $\frac{100c}{2}$ times, where $c$ is the complexity of the task.

\subsubsection{Software Kernel Architecture}

The software kernels are written in OpenCL and are identical for CPU and GPU. They consist of a parallelizable part where each work item is assigned to the computation of exactly one output value $r_i$ and a serial part where all values of the output array are processed by the first work item. The total complexity of the task is divided into the two parts according to the parallelizability of the task. Since streamability is not relevant for either of the devices, it has no representation in the OpenCL kernel.

\subsubsection{FPGA Kernel Architecture}
\label{ssc:fpgaarchitecture}

The FPGA kernel architecture follows the same general principle as the software kernel architecture, but implements additional measures in order to enable streaming inside a task as well as between FPGA tasks. The FPGA kernel is based on Vitis HLS and must undergo a bitstream generation before it can be integrated into the final heterogeneous system.

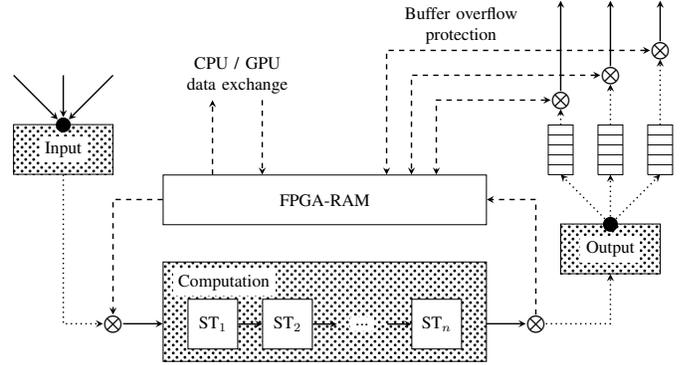
\begin{figure}[h!]
    \centering
    \resizebox{\linewidth}{!}{\begin{tikzpicture}
    \node [draw, pattern={crosshatch dots}, anchor = north west, inner sep = 0mm, minimum width = 6.5cm, minimum height = 2cm]  (Kernel) at (-0.5,0.75) {};
    \node [anchor = north west, inner sep = 0mm, minimum width = 1cm, minimum height = 0.5cm] (COMP) at (-0.3,0.6) {\colorbox{white}{Computation}};
    \node [draw, fill = white, anchor = north west, inner sep = 0mm, minimum width = 1cm, minimum height = 1cm] (A) at (0,0) {ST$_1$};
    \node [draw, fill = white, anchor = north west, inner sep = 0mm, minimum width = 1cm, minimum height = 1cm] (B) at (1.5,0) {ST$_2$};
    \node [anchor = north west, inner sep = 0mm, minimum width = 1cm, minimum height = 1cm] (C) at (3,0) {\colorbox{white}{...}};
    \node [draw, fill = white, anchor = north west, inner sep = 0mm, minimum width = 1cm, minimum height = 1cm] (D) at (4.5,0) {ST$_n$};

    \draw [thick, -stealth] (A) -- (B);
    \draw [thick, -stealth] (B) -- (C);
    \draw [thick, -stealth] (C) -- (D);

    \node [anchor = center, inner sep = 0mm] (InputXOR) at (-1.5, -0.5) {$\bigotimes$};
    \node [anchor = center, inner sep = 0mm] (OutputXOR) at (7, -0.5) {$\bigotimes$};

    \draw [thick, -stealth] (InputXOR) -- (-0.5,-0.5);
    \draw [thick, -stealth] (6,-0.5) -- (OutputXOR);

    \node [draw, anchor = north west, inner sep = 0mm, minimum width = 6.5cm, minimum height = 1cm]  (RAM) at (-0.5,2.5) {FPGA-RAM};

    \draw [thick, -stealth, dashed] (RAM) -- (-1.5,2) -- (InputXOR);
    \draw [thick, -stealth, dashed] (OutputXOR) -- (7, 2) -- (RAM);

    \node [draw, pattern={crosshatch dots}, anchor = south, inner sep = 0mm, minimum width = 2cm, minimum height = 1cm] (I) at (-2.5,2.5) {\colorbox{white}{Input}};
    \node [fill, circle, anchor = center, inner sep = 0mm, minimum size = 3mm] (In) at (-2.5, 3.5) {};
    
    \draw [thick, -stealth, dotted] (I) -- (-2.5, -0.5) -- (InputXOR);


    \node [draw, anchor = south, inner sep = 0mm, minimum width = 2cm, minimum height = 1cm, pattern={crosshatch dots}] (O) at (8.5,0.5) {\colorbox{white}{Output}};
     \node [fill, circle, anchor = center, inner sep = 0mm, minimum size = 3mm] (Out) at (8.5, 1.5) {};

     \draw [thick, -stealth, dotted] (OutputXOR) -- (8.5, -0.5) -- (O);

    \draw [thick, -stealth] (-2.5, 4.5) -- (In);
    \draw [thick, -stealth] (-3.5, 4.5) -- (In);
    \draw [thick, -stealth] (-1.5, 4.5) -- (In);

    \draw [thick, -stealth, dotted] (Out) -- (8.5, 2.5);
    \draw [thick, -stealth, dotted] (Out) -- (9.5, 2.5);
    \draw [thick, -stealth, dotted] (Out) -- (7.5, 2.5);

    \draw (8.25, 2.5) -- (8.75, 2.5) -- (8.75, 3.5) -- (8.25, 3.5) -- (8.25, 2.5);
    \draw (8.25, 2.7) -- (8.75, 2.7);
    \draw (8.25, 2.9) -- (8.75, 2.9);
    \draw (8.25, 3.1) -- (8.75, 3.1);
    \draw (8.25, 3.3) -- (8.75, 3.3);

    \draw (7.25, 2.5) -- (7.75, 2.5) -- (7.75, 3.5) -- (7.25, 3.5) -- (7.25, 2.5);
    \draw (7.25, 2.7) -- (7.75, 2.7);
    \draw (7.25, 2.9) -- (7.75, 2.9);
    \draw (7.25, 3.1) -- (7.75, 3.1);
    \draw (7.25, 3.3) -- (7.75, 3.3);

    \draw (9.25, 2.5) -- (9.75, 2.5) -- (9.75, 3.5) -- (9.25, 3.5) -- (9.25, 2.5);
    \draw (9.25, 2.7) -- (9.75, 2.7);
    \draw (9.25, 2.9) -- (9.75, 2.9);
    \draw (9.25, 3.1) -- (9.75, 3.1);
    \draw (9.25, 3.3) -- (9.75, 3.3);

    \node [anchor = center, inner sep = 0mm] (InputXOR_1) at (7.5, 4) {$\bigotimes$};
    \node [anchor = center, inner sep = 0mm] (InputXOR_2) at (8.5, 4.5) {$\bigotimes$};
    \node [anchor = center, inner sep = 0mm] (InputXOR_3) at (9.5, 5) {$\bigotimes$};

    \draw [-stealth, thick] (InputXOR_1) -- (7.5, 6);
    \draw [-stealth, thick] (InputXOR_2) -- (8.5, 6);
    \draw [-stealth, thick] (InputXOR_3) -- (9.5, 6);

    \draw [thick, dashed, stealth-stealth] (InputXOR_1) -- (5,4) -- (5, 2.5);
    \draw [thick, dashed, stealth-stealth] (InputXOR_2) -- (4.5,4.5) -- (4.5, 2.5);
    \draw [thick, dashed, stealth-stealth] (InputXOR_3) -- (4,5) -- (4, 2.5);

    \draw [thick, dotted, -stealth] (7.5, 3.5) -- (InputXOR_1);
    \draw [thick, dotted, -stealth] (8.5, 3.5) -- (InputXOR_2);
    \draw [thick, dotted, -stealth] (9.5, 3.5) -- (InputXOR_3);

    \draw [dashed, thick, -stealth] (0.5, 2.5) -- (0.5, 4);
    \draw [dashed, thick, -stealth] (1.5, 4) -- (1.5, 2.5);

    \node [anchor = south, align = center] at (1,4) {CPU / GPU\\ data exchange};
    \node [anchor = south, align = center] at (5.5,5) {Buffer overflow\\ protection};
\end{tikzpicture}}
    \caption{The architecture of an FPGA kernel. In its core, the computation is divided into multiple subtasks according to its streamability. Data exchange with CPU and GPU is done using the RAM, whereas communication with other FPGA kernels is based on AXI streams with FIFO buffers.}
    \label{fig:fpga_architecture}
\end{figure}

The architecture of an FPGA kernel is shown in Figure \ref{fig:fpga_architecture}. It consists of a central computation kernel and optional input and output kernels for the communication with other FPGA kernels.
The computation kernel consists of multiple subtask blocks (ST), among which the workload is evenly distributed. The number of subtask blocks corresponds to the streamability parameter. Between each subtask block, a FIFO buffer (sized for a single data element) enables seamless streaming of data through the pipeline. The subtask blocks operate as a dataflow and are interlocked, forming a pipeline that can significantly reduce the execution time.

\begin{figure}[htb]
    \centering
    \includegraphics[width=\linewidth]{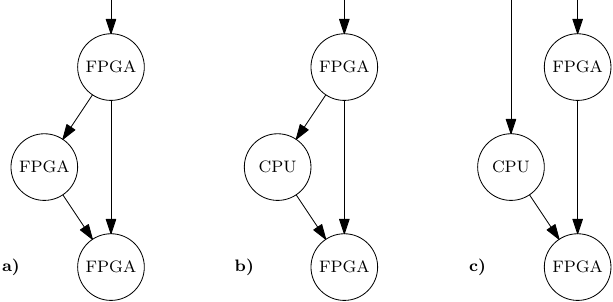}
    \caption{Three different cases that complicate streaming between tasks. An interposed FPGA node delays the stream, whereas an interposed CPU node makes streaming impossible. An otherwise unrelated incoming CPU node may or may not hinder the streaming depending on its finish time.}
    \label{fig:buffer_cases}
\end{figure}

For streaming data to or from another FPGA kernel, an additional output or input kernel is generated that distributes or collects the streaming data via AXI stream, respectively. The output kernel uses FIFO buffers in order to bridge delays between kernel execution times. If the buffer is full before the next kernel starts to consume data, it is expected that the kernel cannot be executed or the delay is too high for streaming to be efficient and the output is written into the RAM instead.  In this work, we use a FIFO buffer of size $20$, which is sufficient for all considered test cases.
Note that it is non-trivial to determine statically whether streaming between two FPGA kernels should be employed. Figure~\ref{fig:buffer_cases} depicts three cases in which (a) streaming can and should be done by using appropriate buffers, (b) streaming is not possible and (c) streaming could be done but may be detrimental based on runtime parameters, i.e., based on the finishing time of the depicted CPU task.

\subsection{Heterogeneous integration and evaluation}

For the final evaluation, an OpenCL host program integrates the CPU/GPU kernels and the FPGA bitstream into a single heterogeneous system based on the dependency structure given by the task graph. Task are executed in separate threads and scheduled dynamically using OpenCL events. A task is executed as soon as all of its input data is available, except in the case of FPGA to FPGA streaming, where a kernel is scheduled simultaneously with its parent kernels as long as all non-FPGA dependencies are fulfilled.

\section{Evaluation}
\label{sec:evaluation}

In this section, we demonstrate the capabilities of the presented framework by evaluating the rapid makespan prediction model presented in~\cite{wilhelm2023}. We generate random graphs and map them using a simulated annealing approach, which itself uses the rapid makespan prediction function as the optimization criterion. We evaluate both pure CPU-GPU mappings and mappings involving FPGA kernels.

\subsection{System configuration}

The test system consists of an AMD Epyc 7351P CPU with 16 cores, an AMD Radeon Vega 56 graphics card and a VCK5000 Versal development card for the programmable logic.
In our experiments, the components are used both as a reference for the platform model needed for the makespan prediction and to run the evaluation platform and execute the actual kernels. Through fast profiling, we deduced that each CPU core is able to process about $11.3$ GB/s, whereas each GPU core can process about $2.4$ GB/s. For parallelizable tasks, the CPU can distribute the work to up to $16$ cores while the GPU can use up to $3584$ cores. The streaming rate of the FPGA board is at about $1.6$ GB/s. Currently, we do not explicitly simulate the area usage of FPGA kernels. Instead, we use an arbitrary constraint of $64$ area units, whereas the amount of area units a task consumes is set to be equal to its streamability. For all tests, we use randomly generated series-parallel graphs (cf.~\cite{wilhelm2023}) and process 10 MB of data.

\subsection{Mapping approach}

In this work, we use both pure CPU mappings and mappings determined by a simulated annealing algorithm. Simulated Annealing is one of the most frequently used metaheuristics in the context of task mapping. We designed a simulated annealing algorithm according to the guidelines presented by Orsila et al.~\cite{orsila2013}. We use a single task move function with a geometric temperature schedule ($q=0.95$) and a normalized exponential acceptance function. The initial and final temperature are chosen systematically based on the ratio between minimum and maximum execution time of the tasks on the devices ($k=2$, see~\cite{orsila2009} for details). We take the best result out of $10$ runs and use $50$ iterations per temperature level. Although Orsila et al.\ recommend to scale the iterations with the graph size, we did not notice a significant difference when using more iterations and hence use a constant number in order to decrease the execution time. In each iteration, the quality of a single design point is determined by using rapid makespan prediction with an assumed breadth-first schedule~\cite{wilhelm2023}.

\subsection{CPU-GPU mappings}

In a pure CPU-GPU system, the presented framework is able to generate and evaluate a large amount of test graphs of arbitrary size in a short time. For the evaluation of the makespan prediction, we generate $100$ graphs with $50$ nodes each and map them using simulated annealing. For graphs of this size, the kernel generation takes about \SI{5}{\milli\second} and the host code execution has an overhead of about \SI{3}{\second}. The optimized mapping improves on the pure CPU mapping in 92 cases by up to \SI{24}{\percent} with an average makespan reduction of \SI{8}{\percent}. While many tasks can be accelerated using a GPU, this often just leads to a shift in the critical path towards a path dominated by tasks that cannot be effectively parallelized and hence are mapped to the CPU. Therefore, the optimization potential using only GPUs is limited for the chosen task graphs.
The makespan prediction is highly reliable, with an average makespan error of about \SI{5}{\percent} for the basic mapping and \SI{4}{\percent} for the optimized mapping. These results suggest that, in this scenario, a rapid makespan prediction is sufficient for an early design space exploration using a task mapping algorithm.

\subsection{CPU-GPU-FPGA mappings}

For a platform containing one or more FPGAs, the validation of a prediction is bottlenecked by the generation of the HLS bitstream. While the prediction itself takes only a few milliseconds to compute, generating a bitstream for the generated FPGA kernels takes about two to three hours to complete. Hence, we use a lower number of smaller graphs for the evaluation compared to the previous scenario. 

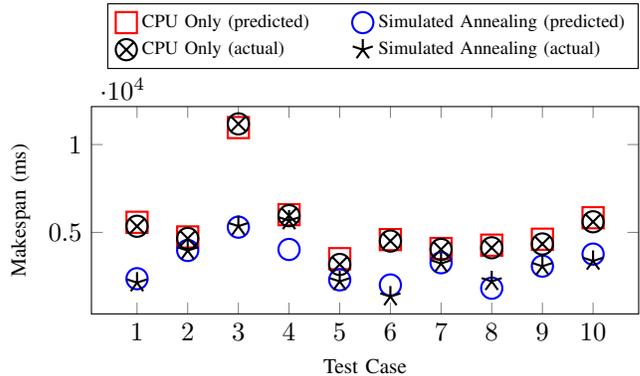
\begin{figure}[ht]
\centering
\begin{tikzpicture}
\begin{axis}[only marks,xlabel={Test Case}, legend columns=2, legend to name=fpgalegend, height=0.49\linewidth,width=\linewidth,legend cell align={left},legend style={font=\scriptsize, /tikz/every even column/.append style={column sep=0.5cm}},label style={font=\footnotesize},ylabel={Makespan (ms)},every axis plot/.append style={thick,mark options={scale=2}},xtick distance = 1]

\addlegendentry{CPU Only (predicted)}
\addlegendentry{Simulated Annealing (predicted)}
\addlegendentry{CPU Only (actual)}
\addlegendentry{Simulated Annealing (actual)}

\addplot[color=red, mark=square] coordinates{(1,5574) (2,4742) (3,10967) (4,6009) (5,3505) (6,4590) (7,4084) (8,4274) (9,4612) (10,5834) };

\addplot[color=blue,mark=o] coordinates{(1,2364) (2,3971) (3,5299) (4,4028) (5,2299) (6,2007) (7,3273) (8,1823) (9,3080) (10,3764) };

\addplot[mark=otimes] coordinates{(1,5349) (2,4655) (3,11178) (4,5951) (5,3175) (6,4507) (7,4030) (8,4125) (9,4341) (10,5604) };

\addplot[mark=star] coordinates{(1,2140) (2,3965) (3,5356) (4,5672) (5,2210) (6,1344) (7,3229) (8,2212) (9,3072) (10,3373) };

\end{axis}
\node[anchor=south west] at ($(current bounding box.north west)!.15!(current bounding box.north east)$) {\ref{fpgalegend}};
\end{tikzpicture}
\caption{Predicted and actual execution times for a pure CPU mapping and a mapping derived through simulated annealing for ten task graphs of size $20$.}
\label{fig:fpgagraphs}
\end{figure}

 In Figure~\ref{fig:fpgagraphs}, the predicted and actual makespans for $10$ graphs with $20$ nodes each are shown. On average, the simulated annealing mapping algorithm reduces the makespan by about \SI{37}{\percent} with a predicted average relative improvement of about \SI{40}{\percent}. It is evident that the makespan prediction is generally accurate. The average relative error is at about \SI{4}{\percent} for the CPU mapping and at about \SI{12}{\percent} for the mappings found by simulated annealing. The average error for simulated annealing is slightly skewed towards a few outliers with a median relative error of \SI{7}{\percent}. In all cases, the predicted makespan improvement of the simulated annealing mapping transferred into an actual improvement in the real system.
 
\begin{figure}[htb]
    \centering
    \begin{subfigure}[c]{.49\linewidth}
        \includegraphics[width=\linewidth,trim={1cm 1cm 1cm 1cm},clip]{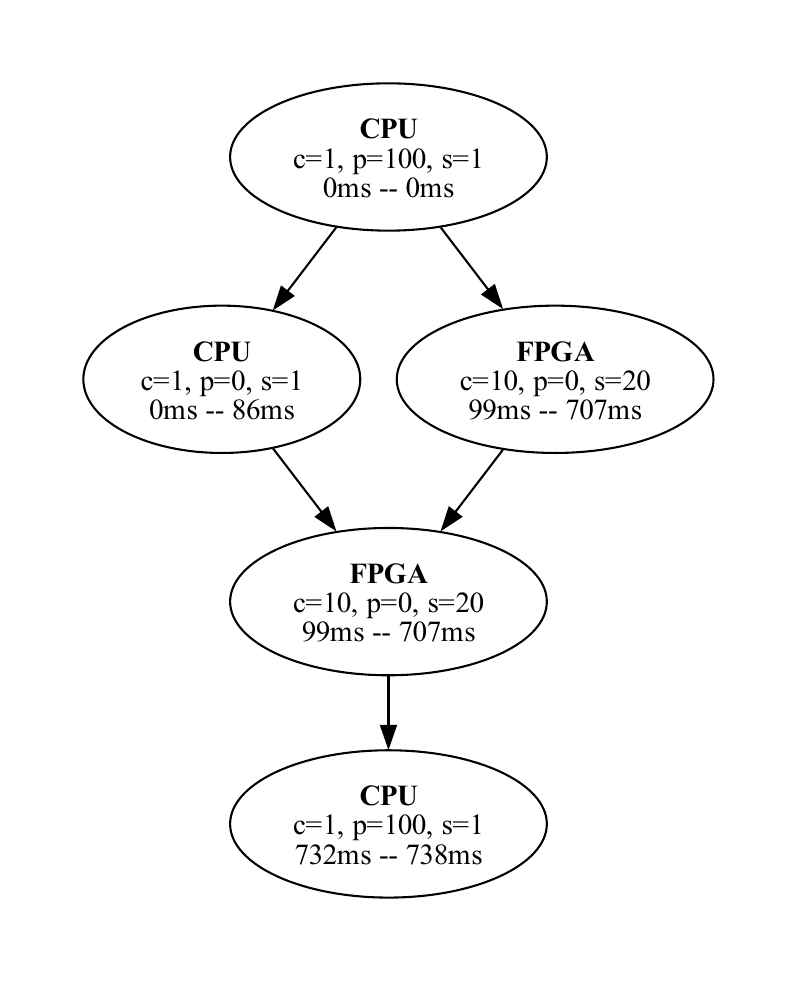}
        \caption{Predicted execution times}
        \label{sfig:model}
    \end{subfigure}
    \begin{subfigure}[c]{.49\linewidth}
        \includegraphics[width=\linewidth,trim={1cm 1cm 1cm 1cm},clip]{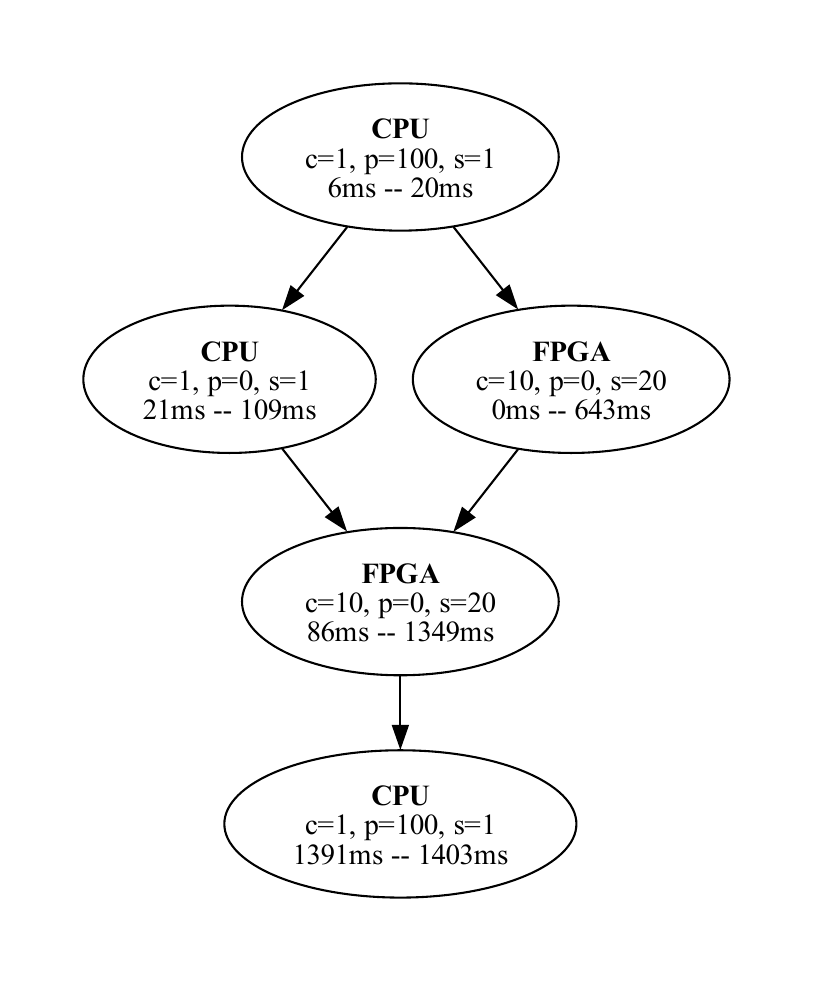}
        \caption{Actual execution times}
        \label{sfig:eval}
    \end{subfigure}
    \caption{Exemplary mapped task graph with predicted and actual execution times, where the decision whether to stream between two FPGA kernels differs. While the prediction model decides to stream between the tasks, the evaluation framework executes them separately. In each node the complexity $c$, parallelizability $p$ and streamability $s$ is depicted together with the time window of the execution. Differences in the start nodes result from an initialization overhead of the real system.}
    \label{fig:fpgatooearly}
\end{figure}	
 In test case~4, the improvement was significantly smaller than expected.
 This discrepancy results from a situation where the real system provided by our framework misjudges whether streaming should be happening between two subsequent FPGA kernels or not (cf.\ Figure~\ref{fig:buffer_cases}). In particular, the system misjudges whether a parent kernel should wait for the child kernel to start working in order to enable streaming or instead fall back to the RAM for the data transfer.

 In Figure~\ref{fig:fpgatooearly}, the predicted makespan and the actual makespan are shown for an exemplary (sub-)graph where the kernel makes a suboptimal decision at runtime. In the example, a streaming connection can be established between the two tasks that are mapped on the FPGA. While the first FPGA kernel can start the computation right away, the second kernel has to wait for the CPU to finish. The prediction algorithm tries to stream between FPGA kernels whenever possible and therefore decides that the first kernel should wait until the second kernel is ready.
 The evaluation framework starts each kernel as soon as possible and decides dynamically whether it makes sense to wait (see Section~\ref{ssc:fpgaarchitecture}). In this case, this leads to a separation between the FPGA kernels. With this decision, the task can be executed in parallel to the CPU task. Since, in this case, the CPU task is executed much faster than the second FPGA task, the approach chosen by the prediction algorithm leads to a lower overall makespan. 
 
 In test case~6, the actual makespan of the optimized mapping is lower than the predicted mapping. Here, two sets of tasks in independent subgraphs are mapped to the FPGA. In particular, the communication with the CPU of one task set overlaps with the computation of the other task set. The used prediction model assumes that the data transfer to a device is delayed if a computation is happening at the device at the same time. While this is generally true for software processors, the assumption does not directly transfer to dataflow processors as long as the current computation does not involve RAM access. In test case~6, this leads to an overestimation of the makespan, since the delayed task is part of the critical path of the application.

\section{Limitations}
\label{sec:limitations}

In the current framework, the produced dummy tasks vary in three key characteristics, their complexity, parallelizability and streamability. With the framework, the behavior of these tasks when mapped onto a heterogeneous platform can be observed and analyzed. With this, the influence of these characteristics, the influence of data transfer between tasks and the influence of device congestion can be analyzed based on an arbitrarily large number of randomly generated applications. In particular, the framework enables the analysis of scaling effects occurring at larger graph sizes. Nevertheless, in the current approach, the executed operations are homogeneous among all tasks. While the dummy operations are sufficient to evaluate whether high-level execution characteristics, i.e., the interaction between different tasks, are sufficiently reflected in a makespan prediction algorithm, it cannot evaluate whether low-level characteristics, i.e., the prediction of individual execution times of tasks on given devices, are done correctly. While, in general, exchanging operations for more expensive ones is simply a matter of linear scaling of the used parameters, some operations might scale differently on different devices. Currently, these base complexities are determined through profiling. In future work, the framework may be extended to use random compositions of different operations, which allows for a more thorough analysis of the prediction of individual execution times.

\section{Conclusion}
\label{sec:summary}

In this work, we present a framework that enables the evaluation of makespan predictions on real heterogeneous systems using randomly generated applications.
The experiments suggest that, in many cases, a rapid prediction is sufficient to compare different task mappings and can generate valuable insights for a design space exploration. With the involvement of streaming processors, dependencies between tasks can lead to a complex decision process during prediction, which must be accounted for by the algorithm. In particular, a change in the order of execution or the granularity of the task graph can make a significant difference for the overall makespan and, hence, for the validity of the prediction.
The three generated task characteristics are able to cover a large portion of the differences between CPUs, GPUs and FPGAs. Based on these characteristics, mapping decisions can be effectively navigated. While this work is focused on the aforementioned devices, we expect these results to be transferable to accelerators such as AI units, which also offer partial support for dataflow processing.

Overall, the incorporation of our generative evaluation framework into the algorithm development cycle is a significant step towards bridging the gap between theory and practice for highly efficient high-level makespan prediction algorithms. While there are still various open problems, e.g.\ with regard to complexity scaling, the framework can effectively capture most high-level effects such as the effect of data transfer, data streaming, execution order and device congestion and, hence, guide algorithm developers towards a better understanding and representation of these aspects in their model.

\bibliographystyle{IEEEtran}
\bibliography{literature}

\end{document}